# Mapping Information in Feature Extraction Transformation for Chirp Signal

Shuyi Gu, Zhenghua Luo*, Lin Hu, Yilin Zhang, and Junxiong Guo*

*Abstract*—Chirp signals have established diverse applications caused by the capable of producing time-dependent linear frequencies. Most feature extraction transformation methods for chirp signals focus on enhancing the performance of transform methods but neglecting the information derived from the transformation process. Consequently, they may fail to fully exploit the information from observations, resulting in decreased performance under conditions of low signal-to-noise ratio (SNR) and limited observations. In this work, we develop a novel post-processing method called "mapping information model" to addressing this challenge. The model establishes a link between the observation space and feature space in feature extraction transform, enabling interference suppression and obtain more accurate information by iteratively resampling and assigning weights in both spaces. Analysis of the iteration process reveals a continual increase in weight of signal samples and a gradual stability in weight of noise samples. The demonstration of the noise suppression in the iteration process and feature enhancement supports the effectiveness of the mapping information model. Furthermore, numerical simulations also affirm the high efficiency of the proposed model by showcasing enhanced signal detection and estimation performances without requiring additional observations. This superior model allows amplifying performance within feature extraction transformation for chirp signal processing under low SNR and limited observation conditions, and opens up new opportunities for areas such as communication, biomedicine, and remote sensing.

*Index Terms*—Chirp signal, Post-processing, Winger Ville distribution, Cross-terms interference, Hough transform, Algorithm, Time-frequency analysis, Signal detection, Parameter estimation.

## I. INTRODUCTION

CHIRP signal is a typical non-stationary signal characterized by a linear variation of frequency over time and are capable of carrying abundant information. Due to these characteristics, diverse applications including radar [1], biological signal processing [2], and geological exploration [3] have reached a high-level maturity by employing chirp signals. High-performance of information acquisition by processing chirp signal is typically depend on detection and estimation theory [4] and feature extraction methods [5]. A significant obstacle encountered in signal processing, including chirp signal processing, is the presence of low signal-to-noise ratio (SNR) [6], which often leads to failure in estimation and detection processes. Consequently, accurately obtaining the carried information within the chirp signal becomes arduous. One potential solution to mitigate the impact of low SNR is to increase the number of observations in both temporal and spatial dimensions. However, this approach has practical limitations imposed by specific application scenarios. To date, how to effectively achieve superior processing performance while dealing with low SNR and limited observations remains a crucial challenge.

The application of estimation methods in statistical signal processing for chirp signal has demonstrated to be effective in numerous studies. For example, cubic phase function (CPF) [7] method models the phase as a cubic polynomial and offers a suboptimal estimation. Additionally, the least squares estimators (LSEs) [8] have been suggested for estimating the frequency rates and amplitude of chirp signal. In a related work [9], an iterative estimation method equivalent to LSE is introduced, treating parameter estimation problem as a nonlinear programming problem. The results indicate that this method closely approximates the theoretical asymptotic variance of LSE. Another approach focuses on estimating the instantaneous frequency function of chirp signals by introducing continuous-time probabilistic models [10]. The model jointly characterizes chirp signals and their instantaneous frequency functions, without requiring prior knowledge of the form of the instantaneous frequency function. Such methods often provide strict closed form expressions of estimation performance, with many claiming optimality or asymptotic optimality. However, once the SNR decreases, these methods often deviate from the Cramér-Rao lower bound (CRLB) [11]. Consequently, the assistance of filtering or other noise reduction techniques become necessary.

Similarly, the time-frequency representation (TFR) [12] is commonly employed as the primary method for analyzing non-stationary signals, to extract target information [13, 14, 15]. Currently, there are various approaches to obtain TFR,

This work was financially supported by the National Natural Science Foundation of China (grant no. 62201096), the Science and Technology Plan of Sichuan Province (grant no. 2023YFS0426) and the Natural Science Foundation of Sichuan Province (grant no. 24NSFSC0409).

Corresponding authors: Zhenghua Luo (luozhenghua@cdu.edu.cn) and Junxiong Guo (guojunxiong@cdu.edu.cn).
Shuyi Gu (gushuyi@stu.cdu.edu.cn), Zhenghua Luo, and Junxiong Guo are with the School of Electronic Information and Electrical Engineering, Chengdu University, Chengdu 610106, China.
Shuyi Gu, Zhenghua Luo, and Lin Hu (hulin@5ritt.com) are with the Fifth Institute of Telecommunication Science and Technology, Chengdu 610036, China and Sichuan Time Frequency Synchronization System and Application Engineering Technology Research Center, Chengdu 610062, China.
Yilin Zhang (zhang_yi_lin@163.com) is with the Chengdu Jinjiang Electronic System Engineering Company, Ltd., Chengdu 610051, China.

including short time Fourier transform (STFT) [12], wavelet transform (WT) [16], Wigner-Ville distribution (WVD) [17], Hilbert-Huang transform (HHT) [18], and other improved methods. Subsequently, the Hough transform (HT) [19] and Radon transform (RT) [20] were proposed to extract chirp signals on TFR, taking advantage of the energy distribution characteristics of chirp signal. One well-known method is the Wigner-Hough transform (WHT) [21], which combines WVD and Hough transform to detect and estimate chirp signals and is equivalent to fractional Fourier transform [22]. Other types of TFR methods can also achieve the same function by combining HT, such as short time Fourier-Hough transform (STFHT) [23], wavelet-Hough transform [24], chirplet-Hough transform (CHT) [25], Teager-Huang-Hough transform (THHT) [26]. In general, the choice of TFR combined with HT depends on the specific characteristics of the scenario to achieve the best output SNR [15, 21, 27]. However, even though combing HT with TFR can achieve a relatively higher SNR, such methods still exhibit threshold effect at low SNR condition [6, 28]. Inspired by these methods, diverse improved algorithms have been demonstrated to enhance the performance of chirp signal processing, including Pseudo Wigner-Hough transform (PWHT) [15], modified Wigner-Hough transform (MWHT) [29], and local polynomial periodogram Wigner-Hough transform (LPPWHT) [30], which aim to address the failure of WHT under low SNR condition. These methods reduce the location of threshold effects by improving the transformation method without additional observations. Other methods focusing on improving TFR, such as the post-processing methods [31], have also been proposed to improve the processing performance of chirp signals under low SNR. For example, Fourier synchrosqueezed-Hough Transform (FSSHT) [32] and wavelet synchrosqueezed-Hough Transform (WSSHT) [33] are able to achieve higher energy concentration and improve the performance of subsequent HT results via rearranging TFR. Additionally, the two-dimensional (2D) mask method [34] is used to suppress interference on TFR and similar. Similar to 2D mask, slice entropy is employed to assign weight to the point in HT feature plane to implement noise suppression and signal feature enhancement [35]. Besides, the neural network [36] extracts and recognizes features in TFR through its unique architecture [37]. These methods of improving in TFR or feature space can effectively improve performance at SNR, but they usually fail to establish a connection between TFR and feature space, leading to limited utilization of observation information and unacceptable performance degradation under low SNR.

When the performance degradation caused by low SNR cannot be avoided, conducting multiple observations at both the temporal and spatial scales achieve better signal processing performance compared to single observations in the spatiotemporal dimension. This provides a promising method for addressing low SNR issues, which has been widely used, such as coherent [38] or incoherent [39] accumulation between multiple observations. However, there is an upper limit to adding observation points at the spatial scale, such as the number of channels in the array antenna, which is limited by application scenarios and costs [40]. In some specific scenes, multiple observations in the time dimension may not be suitable, such as, radar systems require reliable conclusions within a limited time window and limited observed information for target detection [41].

For the reasons mentioned above, it is crucial to focus on how to handle the information contained in a single observation, as it also contributes to reducing the demand for multiple observations. To achieve acceptable performance at lower SNR and extremely minimize observation redundancy, a framework with strong universality and the ability to obtain more information in limited observations should be applied to further improve performance.

Motivated by these factors, this paper presents a novel resampling model called "mapping information model" to establish a connection between the observation space and feature space. The model takes into account the mapping relationship between samples in these two spaces to extract unique information that is distinct from the spatial distributions. Then, a post-processing method based on the model is proposed, which can be widely applied in feature extraction transformation in TFR. This method enhances signal features and suppresses noise through iterative resampling in the two spaces, without requiring any additional observations. It provides an efficient approach for detecting and estimating parameters in low SNR and limited time window, enabling the utilization of limited observation information to achieve improved signal processing results. An illustrative example demonstrating the transformation of feature extraction into the Hough transform is provided, and the convergence results of the post-processing for iterative resampling is analyzed and presented. Finally, several commonly used TFR numerical simulations are conducted to support our claims.

The organization of this paper is as follows: In Section II, we introduce the mapping information model based on the general representation of feature extraction transformation. Subsequently, we incorporate the Hough transform of TFR into the model and develop an iterative method for noise suppression, which applies weights to TFR or feature plane. Section III presents the performances analysis of the proposed noise suppression method, and Section IV employs numerical simulations to validate our analysis. Finally, in Section V, we provide conclusions and discuss future research prospects for the proposed mapping information model.

## II. MAPPING INFORMATION MODEL

### A. Mapping Information in Feature Extraction

In kernel-based feature extraction, mapping function $\phi$ transforms elements from the input space $\Omega$ to the kernel space $\mathrm{H}$, $\phi: \Omega \to \mathrm{H}$. Typically, this involves projecting $\mathrm{H}$ onto a feature parameter space, $\mathcal{L}: \mathrm{H} \to \mathrm{T}$. Given $m$ feature parameters $\boldsymbol{v}$, the feature extraction of a sample vector $\boldsymbol{r} \in \Omega$ can expressed as [42]:

$$\mathcal{L}(\phi(r)) = \begin{bmatrix} \langle v_1, \phi(r) \rangle \\ \langle v_2, \phi(r) \rangle \\ ... \\ \langle v_m, \phi(r) \rangle \end{bmatrix} \quad (1)$$

By further associating the kernel $\phi$ with the feature $v$, we obtain:

$$\phi_v(r) = \{v; \phi(r)\} \quad (2)$$

This indicates the newly defined kernel $\phi_v$ completes the transformation from the input vector to the feature parameter space $\phi_v : \Omega \to T$. The mapping process in equation (1) can be rewritten as follows:

$$\mathcal{L}(\phi(r)) = \phi_v(r) = \int d\phi_v(r) = T(v) \quad (3)$$

For different feature parameter vectors $v$, they are transformed by different input vectors $r$ in equation (3). While it is possible to directly use the entire input space $\Omega$ as input for the mapping $\phi$, there is always a unique sample vector $r \in \Omega$ that serves as the main reason for projecting the input onto the feature space. Therefore, the elements in vector form a set $\mathbb{R}$ related to the feature parameters vector $v_0$ as follows:

$$\mathbb{R}_{v_0} = \{r_i \in r \mid \phi_v(r) = v_0 = \phi_v(\Omega)\} \quad (4)$$

The integral in (3) can be rewritten as:

$$T(v) = \int_{\mathbb{R}_v} d\phi_v(r_i) \quad (5)$$

In discrete cases, we have:

$$T(v) = \sum \phi_v(r_i) \quad (6)$$
$$r_i \in \mathbb{R}_v$$

It is worth noting that kernels with different parameters result in different sets with corresponding samples. The set $\mathbb{R}$ formed during the mapping process is referred to as the Mapping Set (MS) in this paper. Just as different kernels can extract different information, although the mapping set depends on the current kernel, applying operations other than the kernel function to the elements in mapping set can still yield new information. Essentially, this type of information originates from the process of mapping samples onto feature space using $\phi_v$. Hence, we refer to it as Mapping Information (MI).

The mapping information enables us to impose constraints on the distribution characteristics in the feature space after feature extraction transformation. It can be targeted at each point in feature space because the kernel that determines the mapping set depending on the parameters in feature space. It can be used to change the distribution of projections in the feature space, such as enhancing or filtering, while following a certain constraints or criteria function $L$ that is flexible:

$$J_v = L_v(\mathbb{R}) \quad (7)$$

Similarly, whether the feature extraction transformation based on $\phi_v$ is reversible or not, if there is a mapping $\phi_r^{-1} : T \to \Omega$ transforms the feature element to the input space with parameter $r$, a new MS with vector $v$ from $T$ can be composed. $v$ is derived from the linear or nonlinear transformation of $r \in \Omega$ fundamentally, but they may not equal. The set and the MI could be note as:

$$\mathbb{N}_{r_0} = \{v_i \in v \mid \phi_r^{-1}(v) = r_0 = \phi_r^{-1}(T)\} \quad (8)$$

$$C_r = L_r(\mathbb{N}) \quad (9)$$

The sketch of MI model is shown in Fig. 1. We provide a representation of the mapping kernel function of many-for-one between two spaces. If the mapping is many-to-many, we just apply $J$ or $C$ to the multiple projections after mapping. In the sketch, the boundary between the input space and feature space becomes blurred, and the sequence of mapping between them becomes no longer important. The mapping process always generate new information, when the appropriate mapping kernel is selected, as mentioned above and named MI.

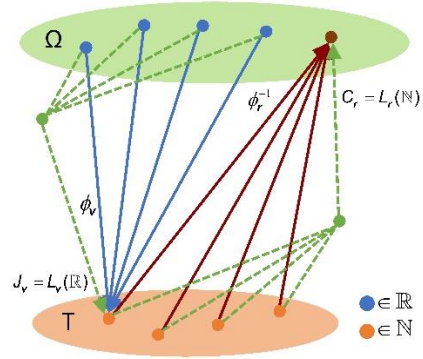

**Fig. 1.** Schematic of Mapping Information model herein.

*B. Weight Matrix based on Mapping Information*

The Hough Transform (HT) is a typical feature extraction transformation or mapping that conforms to the model above described. It is often combined with the TFR of chirp signal. When using the Hough transform for feature extraction, the chirp signal is represented as a straight line in the input space (TFR plane) and an impulse in the feature plane (parameter plane).

The TFR can be defined as:

$$s(t) \to D(t, f) \quad (10)$$

where $s(t)$ is the time domain signal, $D(t, f)$ denotes its TFR. When $s(t)$ is a chirp signal, its energy concentrates along a line on the TFR plane. The Hough transform is then used for detecting the chirp signal and estimating its parameters. In Hough transform, the kernel $\phi$ represents the lines with all possible feature parameters $v = (\rho, \theta)$. These parameters are the polar radius and polar angle of the line in the polar coordinates respectively. In generalized Hough transform (GHT) [43], the kernel can also represent circular shapes and other regular or irregular shapes with different parameters, this makes the discussion solely about HT universal. The mapping kernel of the Hough transform could be written as the form in equation (2):

$$\phi_{\rho, \theta}(D(t, f)) = \{(\rho, \theta); \phi(D(t, f))\} \quad (11)$$

where $\phi$ is

$$\phi(D(t,f)) = \begin{cases} D(t,f) & \rho = t\cos\theta + f\sin\theta \\ 0 & \rho \neq t\cos\theta + f\sin\theta \end{cases} \quad (12)$$

Therefore, following the form of equation (5), the entire feature extraction process can be expressed as:

$$T(\rho,\theta) = \int d\phi_{\rho,\theta}(D(t,f)) \quad (13)$$

For different feature parameters, different mapping sets (MSs) can be generated by kernel (11) in the HT. As discussed above, the resulting MS should be a set that projects onto a specific location in feature space. Here, we consider treating all outputs of $\phi$ as an MS when $\rho = t\cos\theta + f\sin\theta$. Hence, equation (13) can be rewritten as:

$$T(\rho,\theta) = \int_{\mathbb{R}_{\rho,\theta}} dD(t,f) \quad (14)$$

In discrete cases, it is given by:

$$T(\rho,\theta) = \sum_{\mathbb{R}_{\rho,\theta}} D(t_i, f_j) \quad (15)$$

Thus, the MSs formed by the Hough transform are obtained as following.

## C. Using of Mapping Information

When using the HT, we aim to detect a significant impulse in the parameter plane, which indicates the presence of a chirp signal. In the TFR plane, the chirp signal is a line and we assume that its energy is evenly distributed at every point along the line. Therefore, for a chirp signal $s(t)$, we considered it to follow a uniform distribution in the TFR plane, and its random process samples $S$ can be assumed as:

$$S \sim \sqrt{A}U(t_0, t_1) \quad (18)$$

where $A$ is the power of the signal, $t_0$ and $t_1$ represent the start time and end time of the signal, respectively. For additive Gaussian noise $n(t)$ unrelated to the signal, we consider its samples $N$ as a random variable following a zero-mean Gaussian distribution in the TFR plane. These samples are independent, and $\sigma_n^2$ is used to represent the power of the noise:

$$N \sim N(0, \sigma_n^2) \quad (19)$$

Based on the previous discussion, we can see that the mapping

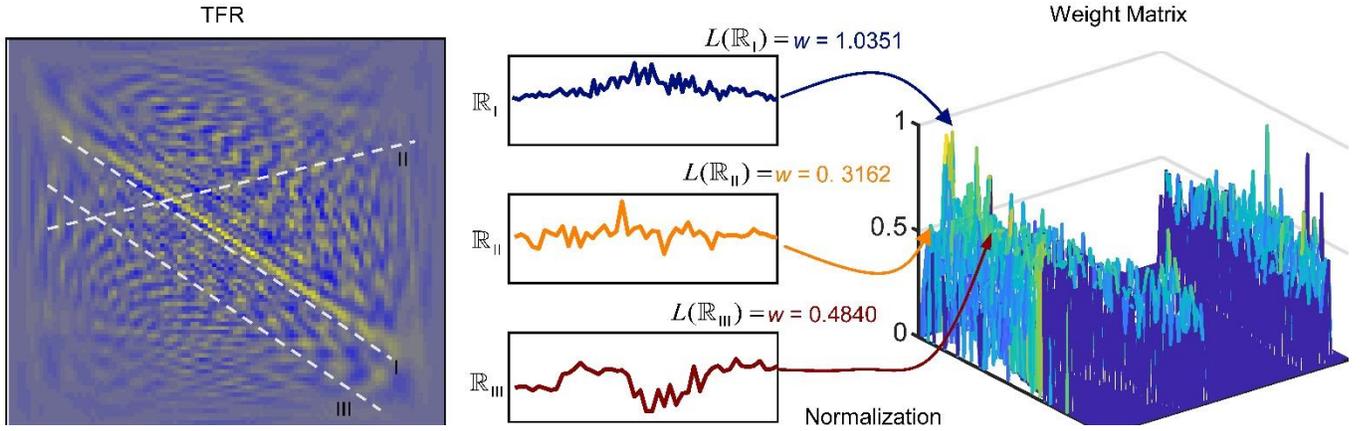

**Fig. 2.** Generate weight matrix from mapping set.

$$D(t_i, f_j) = D_{ij} \in \mathbb{R}_{\rho,\theta} \quad (16)$$

The proposed MS provides information about the transformation process of features extraction from the TFR to the feature parameter space. Each feature $(\rho_0, \theta_0)$ in the feature space corresponds to a set $\mathbb{R}_{\rho_0,\theta_0}$:

$$\begin{aligned} J_{\rho,\theta} &= L_{\rho,\theta}(D_{ij}) \\ D_{ij} &\in \mathbb{R}_{\rho,\theta} \end{aligned} \quad (17)$$

When the TFR $D(t,f)$ is not sparse, there are significant number of non-zero elements in the set, which provides abundant mapping information for the function $L$. This allows us to use more flexible functions to obtain a feature space distribution that aligns better with our expectations. However, when the TFR is sparse, such as Teager-Huang transform (THT) [44], an increase of zero elements in the MS requires careful function selection and may result in decreased adaptability to interference.

sets $\mathbb{R}_{\rho,\theta}$ formed by random variables $S$ and $N$ during the transforming process have extremely different distributions. This statistical characteristic can be used to obtain additional information without conducting additional observations. In this work, we quantify this statistical characteristic by negentropy [45]. Negentropy is used to measure the difference between the sample distribution and the Gaussian distribution using independent component analysis (ICA) [46], which can be given as:

$$L_v(x) = H(x_G) - H(x) \quad (20)$$

where $H(x_G)$ is the entropy of the Gaussian distribution random variables, $H(x)$ is the entropy of the samples. In this paper, we focus on the detection and parameter estimation of chirp signal. Hence, we consider a function with the form of negentropy for the samples in each MS $\mathbb{R}$, defined as follows:

$$J = L_v(x) = H(x_U) - H(x) \quad (21)$$

$H(x_G)$ is the entropy of random variables with a uniform distribution. When the samples in set $\mathbb{R}_{\rho_0,\theta_0}$ are from the chirp signal in the TF plane, it means that they are uniformly distributed. The output value of function is zero (in real life scenarios where there is unavoidable noise, it is a small value). Otherwise, if there are only noise in the mapping set, the distribution is different from uniform distribution and the output of the function have a relatively large value. Furthermore, the output value has a one-to-one correspondence with the parameter plane. Therefore, a weight matrix can be constructed using all outputs of the function:

$$w(\rho,\theta) = J_{\rho,\theta} = L_v(\mathbb{R}_{\rho,\theta}) \tag{22}$$

Here, when we use negentropy as the weight, it can be expressed in the following form:

$$\begin{aligned}w(\rho,\theta) &= H(x_U) - H(x) \\ &= \sum -p_{x_U} \log p_{x_U} - \sum -\tilde{p}_x \log \tilde{p}_x\end{aligned} \tag{23}$$

where $p_{x_U}$ is the probability of uniform distribution, $\tilde{p}_x$ is the frequency about the element in mapping set $\mathbb{R}_{\rho,\theta}$.

For a known distribution, a closed form solution for the entropy of random variable samples can be determined, but it is not so easy for samples in a mapping set. Therefore, we assume that the samples in each mapping set follow a Gaussian distribution when considering the negentropy of each mapping set. Based on this, the weight can be expressed as shown in equation (24), where $l$ is the number of elements in this mapping set, $\tilde{\sigma}_x$ is their standard deviation, and $\mu$ is the mean.

$$J_{\rho,\theta} = H(x_U) - H(x) \Leftrightarrow H(x_U) - H(x)|_{x \sim N(\mu,\tilde{\sigma}_x^2)} \tag{24}$$

that is,

$$\begin{aligned}\sum -p_U \log p_U - \sum -\tilde{p} \log \tilde{p} &\Leftrightarrow \log l - \log \sqrt{2\pi}\tilde{\sigma}_x - \frac{1}{2} \\ &= \log \frac{l}{\sqrt{2\pi}\tilde{\sigma}_x} - \frac{1}{2}\end{aligned} \tag{25}$$

By examining the weight formula provided, it is clear that the weight of each point on the parameter plane depends on the number of samples $l$ and their variance $\tilde{\sigma}_x^2$ in MS $\mathbb{R}$, but is not related to $\mu$. When all the elements within the set are time-frequency (TF) points generated by noise, this assumption is unproblematic. When there is a set with chirp signal TF point, $\tilde{\sigma}_x$ is small, and $l$ is not too small. For an MS composed by chirp signal samples, the term $l/\sqrt{2\pi}\sigma_x$ is larger than when there is a noise set. To better illustrate this, Figure 2 shows three mapping sets generated during feature extraction of TFR using Hough kernel with different parameters. These mapping sets are represented by a line coincident with chirp signal point in the TFR, a line that only passes through noise points in the TFR, and a line that passes through noise points but partially overlaps with a chirp signal point. As the figure shows, these mapping sets have different sample distribution characteristics. By quantifying these characteristics using function $L(\cdot)$, we obtain different weight values. Figure 2 demonstrates that our weight formula can effectively assign different weights to noise and chirp signals in the feature plane, as the logarithmic function is strictly monotonic. To ensure positive weights, we ultimately use the weight calculation given in equation (26).

$$w_{\rho,\theta} = \log(1 + \frac{l}{\sqrt{2\pi}\tilde{\sigma}_x}) \tag{26}$$

After obtaining the corresponding weights for each point, we multiply the normalized weight matrix onto the parameter plane. We simulate chirp signals under noise and use WHT to obtain the original distribution in parameter. Figure 3 shows the contour map of the original distribution and the distribution after using weights. In contrast, the latter distribution contains significantly less noise. As we proposed, due to the addition of mapping information, some information that cannot be obtained in the original distribution is extracted and are used to suppress noise through our designed weight formula. This promising approach paves the way for tasks such as chirp signal detection and parameter estimation in noisy environments, where accurately distinguishing between noise and signal is critical.

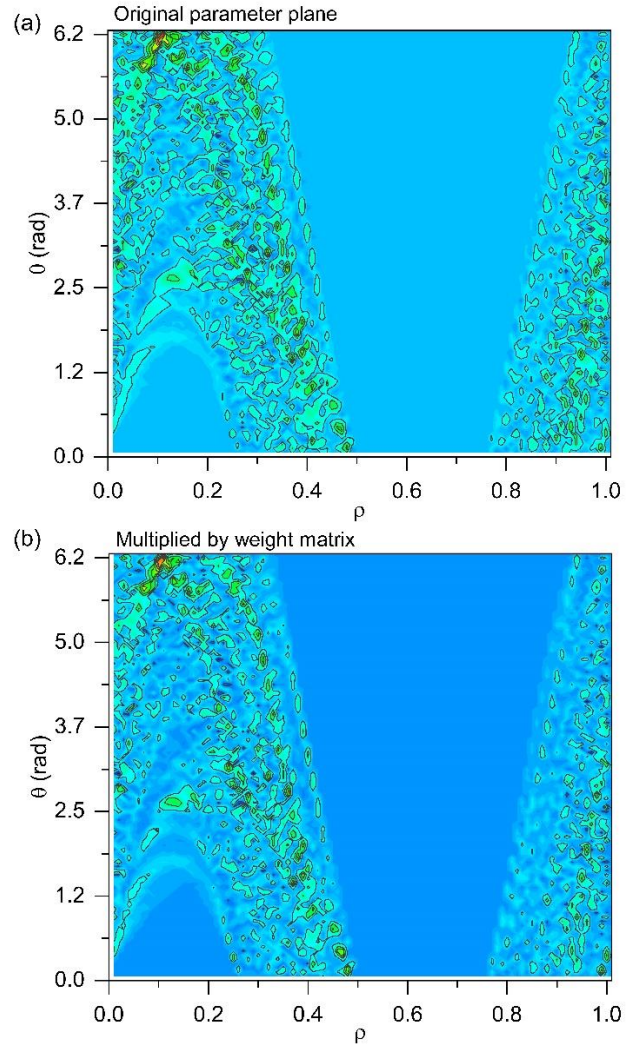

**Fig. 3.** Contour map of (a) original parameter plane distribution and (b) parameter plane distribution after being multiplied by weight.

When mapping from the parameter plane into the TFR plane, the kernel becomes sinusoidal, and we subsequently have followings equations and the generated mapping set $\mathbb{N}$.

$$\phi_{t,f}^{-1}(T(\rho,\theta)) = \{(t,f); \phi^{-1}(T(\rho,\theta))\} \quad (27)$$

$$\phi^{-1}(T(\rho,\theta)) = \begin{cases} T(t,f) & \rho = t\cos\theta + f\sin\theta \\ 0 & \rho \neq t\cos\theta + f\sin\theta \end{cases} \quad (28)$$

As a chirp signal is represented as an impulse on the parameter plane, equation (26) can also serve as a weight on the TFR plane. This is because a prominent impulse increases the variance of the samples within the mapping set. However, as mentioned above, the choice of function $L(\cdot)$ is flexible as long as it adjusts the distribution as desired. To illustrate this point, we introduce another metric called cross-entropy to calculate weights on TFR plane by following equation.

$$\begin{aligned} w(t,f) = C_{t,f} &= L_r(\mathbb{N}_{t,f}) \\ &= H(x_U, x) - H(x, x_U) \quad (29) \\ &= \sum -\frac{1}{l}\log \tilde{p}_x - \sum -\tilde{p}_x \log \frac{1}{l} \end{aligned}$$

Similar to equation (24), we assume that the sample follows a Gaussian distribution:

$$H(x_U, x) - H(x, x_U) \Leftrightarrow H(x_U, x)|_{x \sim N(\mu, \tilde{\sigma}_x^2)} - H(x, x_U)|_{x \sim N(\mu, \tilde{\sigma}_x^2)} \quad (30)$$

Considering the limited number of elements in the mapping set, the weight is given by:

$$\begin{aligned} w_{t,f} &= -\int \frac{1}{l} \log \tilde{p}_x(x) dx + \int \tilde{p}_x(x) \log \frac{1}{l} dx \\ &= \frac{\sqrt{2\pi}\tilde{\sigma}_x}{l} + \log l + \frac{1}{2l\tilde{\sigma}_x^2} \int (x-\mu)^2 dx \quad (31) \\ &\Leftrightarrow \frac{\sqrt{2\pi}\tilde{\sigma}_x}{l} + \log l + \frac{(x_{\max}-\mu)^3 - (x_{\min}-\mu)^3}{6l\tilde{\sigma}_x^2} \end{aligned}$$

Furthermore, it is easy to see that $(x_{\max}-\mu)^3 \geq (x-\mu)^3$ and $(x_{\min}-\mu)^3 < 0$, thus resulting in $(x_{\max}-\mu)^3 - (x_{\min}-\mu)^3 > \tilde{\sigma}_x^3$. Therefore, we have:

$$\frac{(x_{\max}-\mu)^3 - (x_{\min}-\mu)^3}{6l\tilde{\sigma}_x^2} > \frac{\tilde{\sigma}_x^3}{6l\tilde{\sigma}_x^2} = \frac{\tilde{\sigma}_x}{6l} \quad (32)$$

To facilitate the analysis of the weight properties, after shrinking the weight formula as formula (32), we can rewrite the weight calculation formula (31) as:

$$w_{t,f} = \frac{\sqrt{2\pi}\tilde{\sigma}_x}{l} + \log l + \frac{\tilde{\sigma}_x}{6l} \quad (33)$$

It can be observed that the weight of the TFR plane given in this paper depends on the sample variance of the curve corresponding to that point on the parameter plane. The mapping set $\mathbb{N}$ containing impact points have a larger sample variance compared to the set containing only noise points, resulting in a higher weight.

Same way as using weights on parameter plane, we also multiply the weight corresponding to the TFR plane by the original distribution. In our simulations, we generated four chirp signals with different parameters in presence of noise, where the SNR is 3. The time-frequency distribution used is the WVD, and the feature extraction kernel that forms the mapping set remains the Hough transform kernel.

For dealing with multi-component signals, the most well-known drawback of the WVD is cross interference [47]. When the function $L(\cdot)$ is appropriately selected, the proposed mapping information model can handle the cross interference caused by multiple components. We attempt this using (33) and compare the results before and after using weights in Fig. 4. It can be observed that the noise and cross interference in Fig. 4a are significantly suppressed in Fig. 4b, while the chirp signal is enhanced. This is achieved through the high output SNR of the Hough transform feature extraction kernel and the suitable extraction of mapping information for both signal and noise using function $L$. The comparison results in Fig. 4 demonstrate the potential of the mapping information model in cross interference suppression. However, it is worth noting that this result is just an illustration and is not limited to it. The proposed mapping information model allows for flexible selection of functions $L(\cdot)$ to achieve the desired goal. More specific performance analysis of the two weight equations presented in this paper is elaborated in detail in the next section.

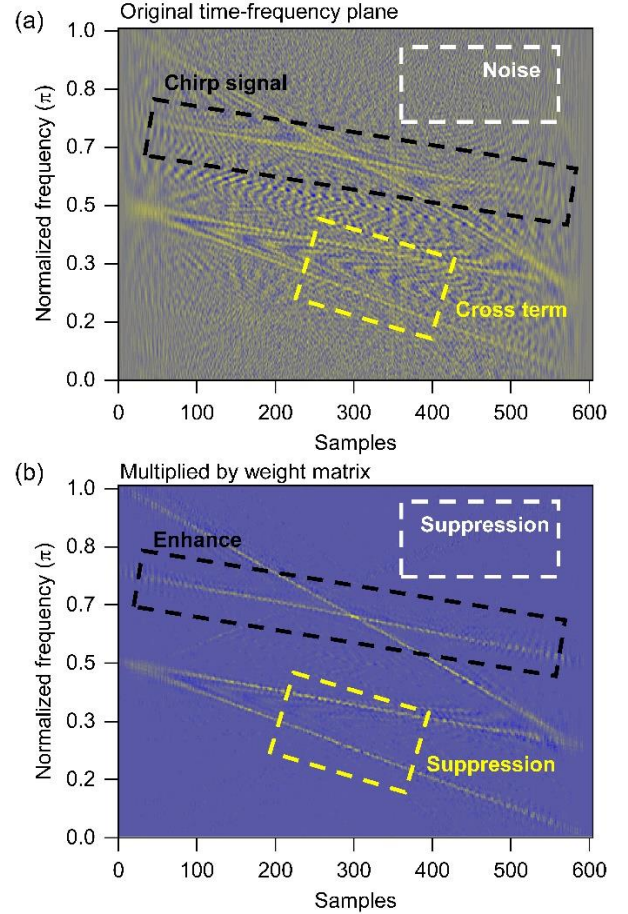

**Fig. 4.** (a) Original time-frequency plane distribution and (b) weighted time-frequency plane distribution.

*D. The Connecting Effect of Mapping Information*

After obtaining the weights using the equation (26) and (33), the approach in this article is to multiply the parameter plane distribution matrix or TFR plane to obtain a new distribution. Another perspective to view the TFR, Wigner distribution is originally used to describe the probability distribution in phase space [48]. Inspired by this, we try to view the TFR and explain the role of weights from a perspective of probability distribution.

Considering the distribution in TFR plane and parameter plane, we assign the values on the parameter plane to a probability note as $P(X = r_{t,f})$, $P(Y = v_{\rho,\theta})$. The probability form of each point on the parameter plane can be written as:

$$P(X = r_{t_i,f_j}) = p_{t,f}(x) = \frac{r_{t_i,f_j}}{\sum r_{t,f}} \quad (34)$$
$$r_{t,f} \in \Omega$$

$$P(Y = v_{\rho_i,\theta_j}) = p_{\rho_i,\theta_j}(y) = \frac{v_{\rho_i,\theta_j}}{\sum v_{\rho,\theta}} \quad (35)$$
$$v_{\rho,\theta} \in \Omega$$

where $r_{t,f}$ and $v_{\rho,\theta}$ are the values of each point in the two planes. Although the Wigner distribution satisfies the normalization property, the probability distribution on other TFRs and the parameter domain may not satisfy this, while it is not important for the actual processing process, as the difference only lies in whether to divide by a constant. Using the probability characteristic of weights, we express the process of multiplying the original distribution by the weight matrix as:

$$\hat{p} = wp \quad (36)$$

As weights can act as new descriptions of distribution probabilities generated based on existing samples and hypothetical constraints, we have:

$$w_{\rho,\theta} = J_v(\mathbb{R}) = \frac{P(\mathbb{R} \mid Y)}{P(\mathbb{R})}$$
$$= \frac{P(X_0 = x_0,...,X_{l-1} = x_{l-1} \mid Y)}{P(X_0 = x_0,...,X_{l-1} = x_{l-1})} \quad (37)$$

$$w_{t,f} = J_r(\mathbb{N}) = \frac{P(\mathbb{N} \mid X)}{P(\mathbb{N})}$$
$$= \frac{P(Y_0 = y_0,...,Y_{l-1} = y_{l-1} \mid X)}{P(Y_0 = y_0,...,Y_{l-1} = y_{l-1})} \quad (38)$$

Then, there is:

$$\hat{p}_{\rho,\theta} = \frac{P(X_0 = x_0,...,X_{l-1} = x_{l-1} \mid Y)P(Y)}{P(X_0 = x_0,...,X_{l-1} = x_{l-1})}$$
$$= P(Y \mid X_0 = x_0,...,X_{l-1} = x_{l-1})$$
$$= P(Y \mid \mathbb{R}) \quad (39)$$

$$\hat{p}_{t,f} = \frac{P(Y_0 = y_0,...,Y_{l-1} = y_{l-1} \mid X)P(X)}{P(Y_0 = y_0,...,Y_{l-1} = y_{l-1})}$$
$$= P(X \mid Y_0 = y_0,...,Y_{l-1} = y_{l-1}) \quad (40)$$
$$= P(X \mid \mathbb{N})$$

This indicates that after normalizing all $w$, it can be seen as a probability distribution. The process of multiplying with weights can be written to the form of Bayesian formula. According to the Bayesian theorem, the new distribution obtained is the correction of the probability distribution in this plane by the mapping set formed through the Hough transformation process. Meanwhile, according to equations (39) and (40), we establish a connection between the two distribution planes (or space) by the mapping information brought by mapping set. This reveals the essence of the method proposed in this paper and provides strong support for us to further generalize the above definition.

III. PERFORMANCE ANALYSIS

Mapping information offers a feasibility of making a single correction to the original distribution. However, in the connection between observation space and feature space established through mapping information, there is more information that can be used. For all points in these planes, the process of using MI for correction can be expressed as a transfer process:

$$W^i_{\rho,\theta}v^i_{\rho,\theta} = v^{i+1}_{\rho,\theta}$$
$$= \begin{bmatrix} w^i_{\rho_0,\theta_0} & ... & 0 \\ ... & ... & ... \\ 0 & ... & w^i_{\rho_{l-1},\theta_{l-1}} \end{bmatrix}_{l \times l} \begin{bmatrix} v^i_{\rho_0,\theta_0} \\ ... \\ v^i_{\rho_{l-1},\theta_{l-1}} \end{bmatrix}_{l \times 1}$$
$$= \begin{bmatrix} J_v(\mathbb{R}^i_{\rho_0,\theta_0}) & ... & 0 \\ ... & ... & ... \\ 0 & ... & J_v(\mathbb{R}^i_{\rho_{l-1},\theta_{l-1}}) \end{bmatrix}_{l \times l} \begin{bmatrix} v^i_{\rho_0,\theta_0} \\ ... \\ v^i_{\rho_{l-1},\theta_{l-1}} \end{bmatrix}_{l \times 1} \quad (41)$$
$$= \begin{bmatrix} v^{i+1}_{\rho_0,\theta_0} \\ ... \\ v^{i+1}_{\rho_{l-1},\theta_{l-1}} \end{bmatrix}_{l \times 1} \rightarrow \begin{bmatrix} \mathbb{N}^{i+1}_{t_0,f_0} \\ ... \\ \mathbb{N}^{i+1}_{t_{l-1},f_{l-1}} \end{bmatrix}$$

$$W^i_{t,f}r^i_{t,f} = r^{i+1}_{t,f}$$
$$= \begin{bmatrix} w^i_{t_0,f_0} & ... & 0 \\ ... & ... & ... \\ 0 & ... & w^i_{t_{l-1},f_{l-1}} \end{bmatrix}_{l \times l} \begin{bmatrix} r^i_{t_0,f_0} \\ ... \\ r^i_{t_{l-1},f_{l-1}} \end{bmatrix}_{l \times 1}$$
$$= \begin{bmatrix} J_r(\mathbb{N}^i_{t_0,f_0}) & ... & 0 \\ ... & ... & ... \\ 0 & ... & J_r(\mathbb{N}^i_{t_{l-1},f_{l-1}}) \end{bmatrix}_{l \times l} \begin{bmatrix} r^i_{t_0,f_0} \\ ... \\ r^i_{t_{l-1},f_{l-1}} \end{bmatrix}_{l \times 1} \quad (42)$$
$$= \begin{bmatrix} r^{i+1}_{t_0,f_0} \\ ... \\ r^{i+1}_{t_{l-1},f_{l-1}} \end{bmatrix}_{l \times 1} \rightarrow \begin{bmatrix} \mathbb{R}^{i+1}_{\rho_0,\theta_0} \\ ... \\ \mathbb{R}^{i+1}_{\rho_{l-1},\theta_{l-1}} \end{bmatrix}$$

where $l$ denotes the number of points in each space. $W$ is a $l \times l$ diagonal matrix, $r$ and $v$ are $l \times 1$ vectors. We use the superscript $i$ to represent *ith* correction of the original distribution via mapping information, and the subscript explains the space to which a vector or scalar belong.

From the mapping set proposed in the previous section, it can be seen that the mapping set corresponding to a TFR plane point is composed of points in the parameter plane, and the mapping set of a parameter plane point is composed of points in the TFR plane. This indicates we can use mapping information to modify the distribution of a plane and further make corrections to the other plane. This two-direction correction brings new mapping information to describe another plane, while using mapping information only in one direction does not further generate new information. The following iteration equations are used to describe this process:

$$\begin{aligned}
&\phi_v(\boldsymbol{r}^i) = \boldsymbol{v}^i \\
&\boldsymbol{r}^i \to \mathbb{R}^i \\
&diag(L_v(\mathbb{R}^i))\boldsymbol{v}^i = \boldsymbol{v}^{i+1} \\
&\phi_r^{-1}(\boldsymbol{v}^i) = \boldsymbol{r}^i \\
&\boldsymbol{v}^{i+1} \to \mathbb{N}^i \\
&diag(L_r(\mathbb{N}^i))\boldsymbol{r}^i = \boldsymbol{r}^{i+1}
\end{aligned} \quad (43)$$

This iterative process does not require additional observation information, only except a set of samples with original distribution. Essentially, it is a resampling process that provides criteria for the original sample according to the function $L(\cdot)$. The bidirectional alternating iteration process brings new information. Meanwhile, as we have emphasized, the choice of descriptive criteria or constraint functions are flexible. Therefore, for the iterative process given by formula (43), we didn't provide detailed closed results, but analyzed the process from the perspective of how the selected function affects the trend of the process, which can better demonstrate the generality of the proposed mapping information model.

Based on previous analysis, the weight of any given point is mainly determined by the sample variance of its corresponding mapping set. A change in weight affects the new distribution trend of each plane. Equation (26) shows the weight of parameter plane points, it depends on the sample standard deviation. Taylor expanding $w$ to the first derivative at $w^i$, there is:

$$w_{\rho,\theta}^{i+1} = w_{\rho,\theta}^i + (\sigma_{\mathbb{R}_{\rho,\theta}}^{i+1} - \sigma_{\mathbb{R}_{\rho,\theta}}^i)\frac{-l}{\sqrt{2\pi}(\sigma_{\mathbb{R}_{\rho,\theta}}^i)^2 + l\sigma_{\mathbb{R}_{\rho,\theta}}^i} \quad (44)$$

Then we have:

$$\Delta w_{\rho,\theta}^i = \Delta \sigma_r^i \frac{-l}{\sqrt{2\pi}(\sigma_r^i)^2 + l\sigma_r^i} \quad (45)$$

Similarly:

$$w_{t,f}^{i+1} = w_{t,f}^i + (\sigma_{\mathbb{N}_{t,f}}^{i+1} - \sigma_{\mathbb{N}_{t,f}}^i)\frac{6\sqrt{2\pi}+1}{6l} \quad (46)$$

$$\Delta w_{t,f}^i = \Delta \sigma_{\mathbb{N}_{t,f}}^i \frac{6\sqrt{2\pi}+1}{6l} \quad (47)$$

Figure 2 illustrates the three scenarios when only one chirp signal is sampled, they denote the samples of set is all signal points or noise points and both signal points or noise points are included, which are all possible scenarios in TFR plane for the mapping set. For the parameter plane, there are only two possibilities, that is the mapping set include the impulse point or not. It is important to note that, when employing the proposed method in this paper, we standardize the new distribution each time the weight is applied to it.

When all elements in the mapping set are noises $\mathbb{R}_{\rho,\theta}^i = \{r_0^n, r_1^n, ..., r_{l-1}^n\}$, due to standardization processing, there is $\lim_{i\to\infty} \sigma_{\mathbb{R}_{\rho,\theta}}^i = 1$, and $\lim_{i\to\infty} \Delta \sigma_{\mathbb{R}_{\rho,\theta}}^i = 0$. This means for points projected entirely by noise, the weight gradually stops changing, that is $\lim_{i\to\infty} \Delta w_{\rho,\theta}^i = 0$. When the set $\mathbb{N}_{t,f}^i = \{v_0^n, v_1^n, ..., v_{l-1}^n\}$ is composed by noise points, the constant weight of noise points in the parameter plane affects the TFR plane simultaneously, which leads $\lim_{i\to\infty} \Delta \sigma_{\mathbb{N}_{t,f}}^i = 0$ and $\lim_{i\to\infty} \Delta w_{t,f}^i = 0$. Conversely, a mapping set $\mathbb{R}_{\rho,\theta}^i = \{r_0^s, r_1^s, ..., r_{l-1}^s\}$ containing the entire chirp signal samples, such a statistical characteristic, is capable of distinguishing itself from noise and obtaining a larger weight. The high weight in the parameter plane also brings a large variance to the mapping set $\mathbb{N}_{t,f}^i = \{v_0^s, v_0^n, ..., v_{l-2}^n\}$. Equation (33) predicts that if the mapping set of a point in the TFR plane includes the point projection by the entire chirp signal samples mapping set, it obtain similar weights as the other points are all noise. This results in a decrease in sample variance, $\lim_{i\to\infty} \sigma_{\mathbb{R}_{\rho,\theta}}^i = 0$, $\Delta \sigma_{\mathbb{R}_{\rho,\theta}}^i > 0$, and a continuous increase in weight, $\lim_{i\to\infty} w_{\rho,\theta}^i = \infty$. As the weight goes up, $v_0^s$ increases proportionally, resulting in an increase in $\Delta \sigma_{\mathbb{N}_{t,f}}^i$. Ultimately, this also leads to a change in the weight of signal points in the TFR plane $\lim_{i\to\infty} w_{t,f}^i = \infty$.

Finally, based on the above discussions, the change in the weight of the mapping set $\mathbb{R}_{\rho,\theta}^i = \{r_0^s, ..., r_{m-1}^s, r_0^n, ..., r_{l-m-1}^n\}$ that includes both noise and chirp signal samples are easier to analyze. Due to the weight of noise points remains a constant, while the weight of the points mapped by the signal continues to increase. In addition to the influence of standardization, the sample variance of the mapping set $\sigma_{\mathbb{R}_{\rho,\theta}}^i$ continues to increase, as noise points become smaller under standardization, while signal points become stronger. That means:

$$\lim_{i\to\infty} \sigma_{\mathbb{R}_{\rho,\theta}}^i = \infty \quad (48)$$

$$\lim_{i\to\infty} \frac{-l}{\sqrt{2\pi}(\sigma_{\mathbb{R}_{\rho,\theta}}^i)^2 + l\sigma_{\mathbb{R}_{\rho,\theta}}^i} = 0 \quad (49)$$

And due to the square term of $\sigma_{\mathbb{R}_{\rho,\theta}}^i$:

$$\begin{aligned}
\lim_{i\to\infty} \Delta w_{\rho,\theta}^i &= \lim_{i\to\infty} \Delta \sigma_{\mathbb{R}_{\rho,\theta}}^i \frac{-l}{\sqrt{2\pi}(\sigma_{\mathbb{R}_{\rho,\theta}}^i)^2 + l\sigma_{\mathbb{R}_{\rho,\theta}}^i} \\
&= \lim_{i\to\infty} \frac{-l(\sigma_{\mathbb{R}_{\rho,\theta}}^{i+1} - \sigma_{\mathbb{R}_{\rho,\theta}}^i)}{\sqrt{2\pi}(\sigma_{\mathbb{R}_{\rho,\theta}}^i)^2 + l\sigma_{\mathbb{R}_{\rho,\theta}}^i} \\
&= \lim_{i\to\infty} \frac{-l(\frac{\sigma_{\mathbb{R}_{\rho,\theta}}^{i+1}}{\sigma_{\mathbb{R}_{\rho,\theta}}^i} - 1)}{\sqrt{2\pi}(\sigma_{\mathbb{R}_{\rho,\theta}}^i) + l} \\
&= 0
\end{aligned} \quad (50)$$

Due to the mapping set $\mathbb{R}_{\rho,\theta}^i$ contains the signal points, the mapping set composed of parameter planes have these two forms $\mathbb{N}_{t,f}^i = \{v_0^n, v_1^n, ..., v_{l-1}^n\}$ and $\mathbb{N}_{t,f}^i = \{v_0^s, v_0^n, ..., v_{l-2}^n\}$. When there are all noise points $v^n$, the weight is applied to the noise point on TFR plane, the same results as the analysis of all noise points MS mentioned above can be obtained. When signal point is included in MS, the weight for signal point is also same as the above analysis of all signal points MS.

Based on the previous discussion, we have compiled the weights for iteratively correcting the original information distribution using the mapping information model. For a point in the parameter plane, if it is a signal impulse, the weight is given by:

$$\lim_{i \to \infty} w^i_{\rho,\theta} = \infty \quad (51)$$

In other cases, equation (52) applies.

$$\lim_{i \to \infty} w^i_{\rho,\theta} = c \quad (52)$$

For the TFR plane, the weight of a signal point is determined by

$$\lim_{i \to \infty} w^i_{t,f} = \infty \quad (53)$$

While the weight of a noise point is given by

$$\lim_{i \to \infty} w^i_{t,f} = c \quad (54)$$

where $c$ represents a constant.

It can be observed that, based on the function $L(\cdot)$ used in this article, the weight remains constant when facing noise points, while it continuously increases when facing signal points. This allows us to achieve noise suppression and signal point feature enhancement through iteration. Figure 5 shows the changes in the weight matrix of the parameter plane in iteration process, where *Order* is the number of iterations in equation (43). The diagram is obtained from the simulation process with an SNR of -5. Initially, during the early iterations, there are no significant impulses in the weight matrix, making it difficult to determine the presence or position of the signal. However, as the iterations progress, the noise is gradually suppressed, leaving a significant impulse at the point where the chirp signal maps to the parameter plane. Similarly, when there is only noise present, the iterative process of weight matrix is almost the opposite. The iteration process is shown in Fig. 6, with the same format as Fig. 5. Under the influence of high intensity noise, there may be obvious impulses in the parameter plane. However, as the iterations progress, these impulses are no longer be retained. Ultimately, the weight matrix becomes chaotic and disorderly, resembling the noise itself, and the weights no longer change. This is consistent with the constant trend of noise point weights mentioned above.

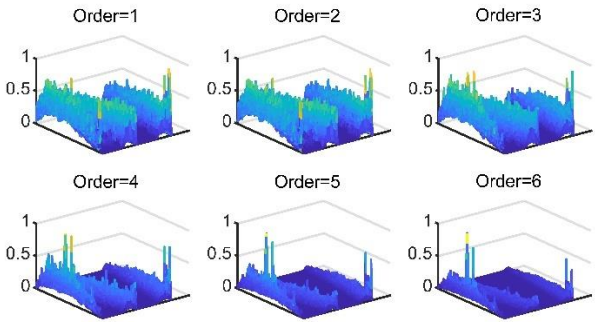

**Fig. 5.** Iterative diagram of parameter plane weight matrix when there is a chirp signal.

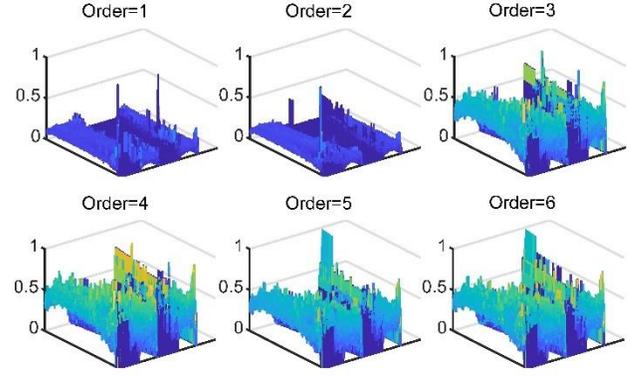

**Fig. 6.** Iterative diagram of parameter plane weight matrix when there is only noise.

IV. SIMULATION RESULTS

*A. Methods*

In this section, we employ the aforementioned iterative process in the simulation to demonstrate the effectiveness of the mapping information model. We focus on the time-frequency distribution of a single chirp signal and its distribution in the Hough transform parameter plane. The time-frequency distributions considered in this work are as follows:

- Wigner-Ville distribution (WVD): It is initially proposed by Wigner [48], demonstrating the joint quantum mechanical distributions of position and momentum. Ville introduced this distribution to signal processing purposes [49]. The WVD exhibits high energy concentration in TF plane, making it suitable for detecting and parameter estimating of chirp signals by combined with the Hough transform. In our results, we represent this method as WHT.
- Fourier synchrosqueezed transform (FSST): The short time Fourier transform (STFT) is a widely-used time-frequency analysis method. To improve time-frequency clustering and approximate the ideal time-frequency representation, the reassignment method (RM) [50] has been proposed as a post-processing technique. The synchrosqueezed transform [51] is a special rearrangement that not only sharpens the time-frequency representation but also facilitates the signal recovery. It offers improved time-frequency resolution and noise robustness. By combining FSST with the Hough transform, that is FSSHT, is introduced for processing chirp signals similar to WHT.
- Wavelet synchrosqueezed transform (WSST): As an equally renowned time-frequency analysis method to STFT, Wavelet transform faces similar challenges. Consequently, it inevitably intersects with the synchrosqueezed transform to achieve higher time-frequency resolution and noise robustness [24]. Similar to FSSHT, we denote this approach as WSSHT;
- Chirplet transform (CT) [52]: The chirplet transform

combines the advantages of the Fourier transform and the wavelet transform, allowing for the utilization of more complex shaped time-frequency atoms to construct TFR. However, it requires the consideration of additional parameters. In this work, we combine the CT and the Hough transform and denote it as CTHT [25].

These four methods for processing chirp signal, WHT, FSSHT, WSSHT, and CTHT, are evaluated in our simulation to demonstrate the effectiveness of the mapping information model.

*B. Evaluation Indicators*

After conducting time-frequency distribution and Hough Transform feature extraction for a chirp signal, we propose to use the following metrics to perform the simulations.

Firstly, detection rate and parameter estimation error are essential metrics in traditional detection and estimation method. In terms of signal detection and parameter estimation, we identify the maximum peak in the parameter plane and compare it with the threshold as shown in equation (55). This is equivalent to GRLT, which has been proven to be an asymptotically optimal detection way for a finite time chirp signal [17]:

$$\max_{\rho,\theta} \int_{-\infty}^{\infty} T(t, \frac{\rho}{\sin\theta} + \tan\theta t) dt \underset{H_0}{\overset{H_1}{\gtrless}} \eta \quad (55)$$

where $\eta$ is the threshold. We use the adaptive threshold [53] following the Neyman-Pearson criterion [4] for a specific false alarm probability:

$$P_f = \int_{\eta}^{\infty} p(x|H_0) dx \quad (56)$$

$$P_d = \int_{\eta}^{\infty} p(x|H_1) dx \quad (57)$$

$P_f$ is false alarm probability, $P_d$ is detection probability, $p(x|H_0)$ and $p(x|H_1)$ are the probability density functions of noise and signal, respectively. When using HT to detect the chirp signal, they are considered as chi-square distributions and non-central chi-square distributions with a degree of freedom of 2.

Secondly, output SNR [27] is widely used to evaluate the performance of the Hough transform and its improved methods. The commonly accepted definition of output SNR is given by [54]:

$$SNR_{out} = \frac{|T_s(\rho_0, \theta_0)|^2}{E\{|T_{s+n}(\rho_0, \theta_0)|^2\}} \quad (58)$$

where $T_s(\rho_0, \theta_0)$ denotes the signal point in parameter plane when there is only signal, $T_{s+n}(\rho_0, \theta_0)$ represents the signal point corrupted by noise. However, considering that the method proposed in this work is not a completely new transform method, but a noise reduction method based on the proposed mapping information. Thus, the impacts caused by the weight matrix should be given more attention. The influence of weight matrix should be taken into consideration when dealing with signals and noises separately. Therefore, we use the following form of output SNR:

$$SNR_{out} = 10\log \frac{|T_{s+n}(\rho_0, \theta_0) - T_n(\rho_0, \theta_0)|^2}{\text{var}\{T_n(\rho_0, \theta_0)\}} \quad (59)$$

Although closed-form expressions for output SNR have been derived in many improved transform methods, the proposed mapping information model may have different forms due to different feature extraction kernels and weight calculation methods. To avoid any misleading results, we did not provide a closed form expression for the output SNR in the analysis. Instead, we used finite iterations of numerical simulation to demonstrate performance. The trend of weight changes was analyzed in the previous chapter, where it was determined that the weight of signal points should be infinite with infinite iterations. However, since the effect of limited iterations is practical, we used numerical simulation to demonstrate performance.

Finally, we introduce a performance metric from side channel analysis (SCA) called Confidence [55] in this study. In SCA, the Confidence is defined as the ratio between the most significant peak and the second significant peak in a one-dimensional sequence. It reflects the credibility of the information represented by the most significant peak. In the context of chirp signal detection, the scene becomes a peak on a two-dimensional plane. Due to the use of adaptive thresholds in signal detection, this work prefers to assess how the detected peak compares to the current threshold. In other words, it measures where the new distributions processed by the method in this article gather above the threshold value. The Confidence metric is used to evaluate this indicator, and the following formula is used to calculate it in following simulation results.

$$Confidence = \frac{T_{s+n}(\rho_0, \theta_0)}{\eta} \quad (60)$$

In the simulations conducted in this paper, a chirp signal with a duration time of $T_r = 10\text{e}-6$ $s$ and a sampling rate of $T_s = 20\text{e}6$ Hz, an initial frequency of $F = F_s/4$, a bandwidth of $B = F_s/4$ and a chirp rate of $G = -B/T_r$ are performed. For the parameters $F$ and $G$ estimation of chirp signal, it is believed that besides parameter estimation during signal detection, the impact of missed or false alarms on parameter estimation should also be considered. Therefore, the principle of minimizing maximum risk in Bayesian decision [56] is emulated, taking into account missed detections in estimation errors. Regarding missed detections, it is assumed that they result in a parameter estimation error of 10% in simulation, as detections with parameter estimation errors greater than 10% are classified as missed detections. The equations used for calculating the estimation errors caused by missed detections or the "risk" brought by missed detections in parameter estimation are as follows:

$$\Delta F = E\{Error\} = \{P_d \frac{|F - \hat{F}_d|}{F} + (1 - P_d) \frac{|F - \hat{F}_m|}{F}\} \times 100\% \quad (61)$$

$$\Delta G = E\{Error\} = \{P_d \frac{|G - \hat{G}_d|}{G} + (1 - P_d) \frac{|G - \hat{G}_m|}{G}\} \times 100\% \quad (62)$$

Here, $\hat{F}_d$ and $\hat{G}_d$ are the estimated values obtained from equation (55), and $\frac{|F - \hat{F}_m|}{F}$ is the error caused by missed detections or the 'risk' associated with missed detections in parameter estimation.

*C. Results*

To evaluate the performance of the proposed method, 1000 Monte Carlo simulations were conducted using different time-frequency distributions combined with the Hough transform under different SNR levels, based on the indicators provided above. When using mapping information in WHT, the significant effect that should be taken into consider is that due to the excellent performance of WVD, it can better conform to the weight calculation formula that we have selected. Figure 7 shows the results of WHT and using mapping information with 1 to 4 iterations on WHT.

After only one iteration, the detection probability has significantly increased, as shown in Fig. 7. Moreover, as the order increases, the improvement in detection probability continues to increase. However, the increment is decreasing, indicating that the performance improvement of continuing to iterate after the weight and distribution of signal points reach their maximum peaks become limited and unnecessary. The output SNR and Confidence exhibiting the same pattern of change, where more iterations lead to greater performance improvement. Additionally, higher orders can bring lower errors for parameter estimation error. Meanwhile, after evaluating the performance using equations (61) and (62), the error curve is not only a numerical indicator of performance, but also an important basis for evaluating when it approaches the set maximum risk value. This is one of the benefits of using them to evaluate parameter estimation errors. As shown in Fig. 7, a higher order leads to an error curve approaching 10% at a lower SNR, which is the 'maximum risk' that has been set. This means that the proposed method has a higher tolerance for noise.

Also, according to the results given in Fig. 7, it can be inferred that the weighting process as a resampling process cannot break through the limitations of its original methods of action. For example, in the Hough transform, there is a threshold effect on performance including detection and parameter estimation, where performance sharply declines when SNR falls below a threshold [28]. This is manifested as various performance evaluation indicators being relatively close at extremely low SNR. Nonetheless, the results in our proposed models still indicate that in some intervals above the threshold SNR, and the proposed method significantly improves various performance of chirp signal processing. Most importantly, our proposed method can essentially be considered as iteratively resampling in two spaces without the need for additional observation information. This enables more efficient utilization of observation information within a limited time window.

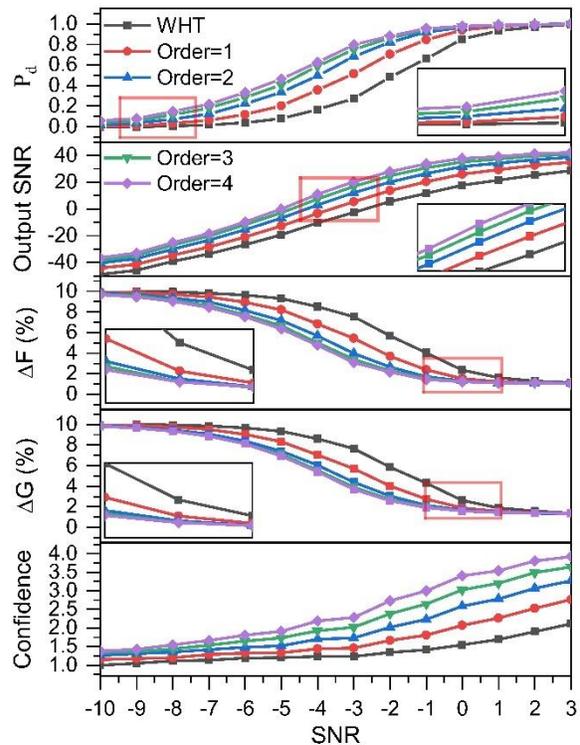

**Fig. 7.** Simulated results of WHT.

Figure 8 shows that the performance of FSSHT is consistent with that of WHT in all aspects. Our proposed method demonstrates a greater improvement in detection probability than WHT, however, it requires a higher SNR range needed. As discussed in Section II, sparser time-frequency distribution reduces the information contained in the mapping set and synchronous compression transformation makes the time-frequency distribution sparse. Furthermore, noise can cause the chirp signal to deviate a straight line in synchronous compression transformation, resulting in a certain degree of mismatch in the weight formula, which affects the performance. Nevertheless, our method still demonstrates improved performance and is consistent with the conclusions of the analysis conclusion.

TABLE I
COMPARISON OF AVERAGE PERFORMANCE

| Metrics / Method | WHT | | | | FSSHT | | | |
|---|---|---|---|---|---|---|---|---|
| | Order=1 | Order=2 | Order=3 | Order=4 | Order=1 | Order=2 | Order=3 | Order=4 |
| $P_d$ | 0.0916 | 0.1568 | 0.1932 | 0.2166 | 0.1106 | 0.2103 | 0.2749 | 0.312 |
| Output SNR | 6.6024 | 11.8401 | 15.2593 | 17.6438 | 3.7362 | 7.454 | 10.2669 | 12.5435 |
| $\Delta F$ | -0.82 | -1.37 | -1.63 | -1.75 | -0.81 | -1.46 | -1.83 | -2.02 |
| $\Delta G$ | -0.75 | -1.24 | -1.42 | -1.47 | -0.6 | -1.01 | -1.19 | -1.27 |
| Confidence | 0.3083 | 0.6129 | 0.8744 | 1.1045 | 0.0874 | 0.2092 | 0.3559 | 0.5035 |
| Metrics / Method | WSSHT | | | | CTHT | | | |
| | Order=1 | Order=2 | Order=3 | Order=4 | Order=1 | Order=2 | Order=3 | Order=4 |
| $P_d$ | 0.0658 | 0.0994 | 0.1211 | 0.1378 | 0.0206 | 0.0374 | 0.0526 | 0.0619 |
| Output SNR | 3.3558 | 5.1953 | 6.5001 | 7.4682 | 1.925 | 3.1588 | 3.9842 | 4.5823 |
| $\Delta F$ | -0.4 | -0.61 | -0.73 | -0.83 | -0.18 | -0.33 | -0.46 | -0.53 |
| $\Delta G$ | -0.2 | -0.3 | -0.34 | -0.38 | -0.16 | -0.29 | -0.4 | -0.46 |
| Confidence | 0.1284 | 0.2278 | 0.3038 | 0.3729 | 0.1224 | 0.1826 | 0.213 | 0.2383 |

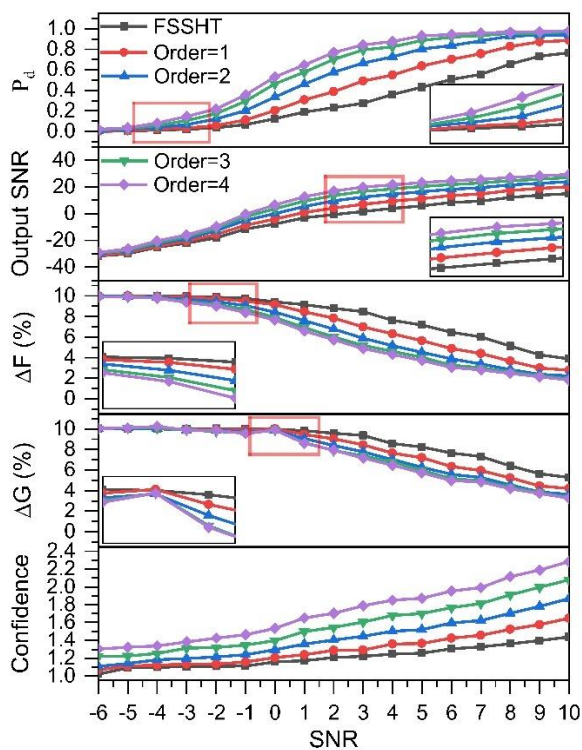

**Fig. 8.** Simulated results of FSSHT.

Similar to FSST, WSST also uses synchronous compression transformation in time-frequency distribution. However, the chirp signal is not a straight line on the time-frequency plane provided by WSST. In Section II, the mapping information model is based on feature extraction transformations that are more general, as opposed to being limited to the Hough transformation. From a general perspective, the mapping information model should be flexible, and the feature extraction kernel and weight calculation formula should be appropriately changed according to the scene to achieve better results. Therefore, we used different feature extraction kernels when processing WSSHT to convert wavelet scale to time-frequency. According to the results given in Fig. 9, our method still shows improved performance with increasing order, even after changing the feature extraction kernel function. Essentially, this is the same as the simulation results of WHT and FSSHT, revealing that the mapping information model is still effective and provides feasible support for the more general applications.

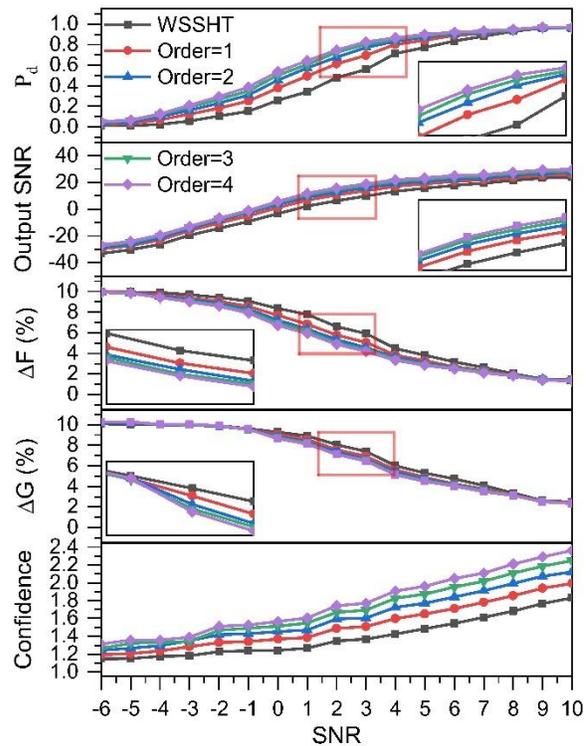

**Fig. 9.** Simulated results of WSSHT.

When applying MI to improve the performance of CTHT, we observed a less significant improvement compared to the other methods, which illustrates importance of selecting appropriate kernel and matching weight calculation formula in mapping information module. As shown in Fig. 10, even after multiple iterations, the performance improvement of CTHT is not substantial. This is mainly because the chirp signal in the time-frequency distribution provided by CT is not strictly a straight line, and selecting appropriate generation function parameters is crucial to ensure the energy concentration on the time-frequency plane [52]. In addition, CT is also susceptible to noise, which can cause distortion of chirp signals in the time-frequency plane. These factors are inconsistent with the feature extraction kernel used in this study and the assumption made when calculating weights based on the time-frequency distribution. Therefore, although increasing the number of iterations can achieve greater performance improvement while maintaining the same trend as the first three methods, CTHT achieves the smallest improvement due to the aforementioned challenges.

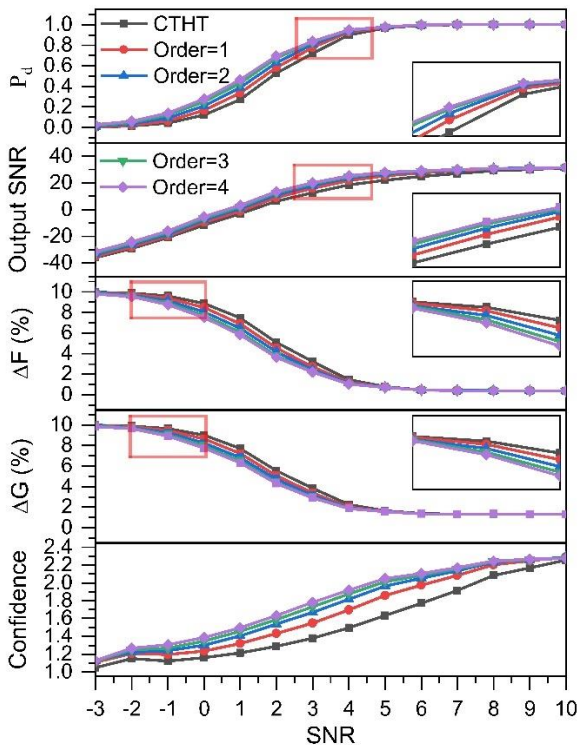

**Fig. 10.** Simulated results of CTHT.

More quantitative information about the average improvement of indicators is summarized in Table I. It can be concluded that an increase in order leads higher detection rate, output SNR, and Confidence. However, as mentioned earlier and depicted in Figs. 7 to 10, there is always an upper limit to this improvement. In terms of parameter estimation, increasing the number of iterations enables it to reach the 'maximum risk' point at a lower SNR, which ensures effective operation even in low SNR conditions. Additionally, for each TFR distribution, the performance improvements observed across different orders highlight the importance of considering appropriate kernel and weight calculation formulas. By doing so greater enhancements in performance can be achieve, thus demonstrating the flexibility inherent in the proposed mapping information model.

## V. CONCLUSION

We have developed a novel model called mapping information model which is utilized in the Hough transform to enhance the performance of chirp signal processing. The demonstrated model enables iterative resampling in both the input space and feature space, thereby altering the distribution of these spaces through generated weights. Moreover, we emphasize that the proposed model offers flexibility and variability in terms of feature extraction kernels and weight calculation. With the proposed model as a basis, we present a weight calculation and iteration method to demonstrate its effectiveness. The simulation results indicate that the proposed model can significantly improve the performance of chirp signal processing at low SNR, with greater improvement achieved through more iterations. Notably, the iterative process does not require additional observation information, making it highly suitable for signal processing devices that need to respond quickly within a limited time window. Furthermore, it is important to note that the proposed model is not limited to the scenarios discussed in this paper. Exploring the selection of more suitable weight calculation and kernel choices based on the proposed model in diversity of scenarios warrants further investigation.